# Synthesis of tunable SnS-TaS$_2$ nanoscale superlattices

*Dennice M. Roberts[1], Dylan Bardgett[1], John D. Perkins[1]†, Brian P. Gorman[2,3], Andriy Zakutayev[1], Sage R. Bauers[1]\**

[1]Materials Science Center, National Renewable Energy Laboratory, Golden CO 80401

[2]Department of Metallurgical and Materials Engineering, Colorado School of Mines, Golden CO 80401

[3]Microelectronics Technology Department, The Aerospace Corporation, El Segundo, CA

KEYWORDS heterostructure, self-assembly, 2D transition metal dichalcogenides

ABSTRACT Nanoscale superlattices represent a compelling platform for designed materials as the specific identity and spatial arrangement of constituent layers can lead to tunable properties. A number of kinetically-stabilized layered chalcogenide nanocomposites have taken inspiration from misfit compounds, a thermodynamically stable class of materials formed of van der Waals-bonded (vdW) layers. This class of vdW heterostructure superlattices have been reported in telluride and selenide chemistries, but have not yet been extended to sulfides. Here we present SnS-TaS$_2$ nanoscale superlattices with tunable layer architecture. Thin films are prepared from layered amorphous precursors and deposited to mimic the targeted superlattice; subsequent low temperature annealing activates self-assembly into designed nanocomposites. Structure and composition for materials are investigated that span stacking sequences between $[(SnS)_{1+\delta}]_3(TaS_2)_1$ and $(SnS)_7(TaS_2)_1$ using x-ray diffraction, x-ray fluorescence, and transmission electron microscopy. A graded deposition approach is implemented to stabilize heterostructures of multiple stacking sequences with a single preparation. Precise control over the architecture of such nanoscale superlattices is a critical path towards controlling the properties of quantum materials and constituent devices.

## INTRODUCTION

Misfit layer compounds (MLCs) exhibit a host of exotic properties making them attractive for basic scientific exploration and for technological applications. Both sulfur- and selenium-based MLCs with PbX monochalcogenide layers, including $[(PbSe)_{0.99}]_m(WSe_2)_n$, have shown ultralow thermal conductivity as a result of incommensurate interfaces, serving as good candidate material for thermoelectric energy conversion technologies.[1,2] Other systems, such as $[(BiSe)_{1+\delta}]_1(NbSe_2)$ exhibit correlated behaviors like superconductivity and charge density localizations,[2,3] where coupling strength is either shown or proposed to be dependent on layer identity and thickness.[4–6] Electrical measurements of $[(PbSe)_{1.16}]_1(TiSe_2)_n$ compounds have unique and dramatic changes to charge transfer effects as $n$ increases from 1 to 2, making it of interest as a material for next-generation computing applications.[7]

Heterostructures formed of layered two-dimensional materials are able to utilize a wide range of novel properties resulting from both the reduced dimensionality of constituent layers and the unique interactions between layers.[8,9] As a subset of this material class, MLCs are naturally-occurring heterostructures, denoted as $[(MX)_{(1+\delta)}]_m(TX_2)_n$ to reflect their structural stacking sequence (ie. $n$ $TX_2$ layers per $m$ $MX$ layer with an areal atom density ratio between lattices of $1+\delta$). In MLCs, TX$_2$ layers are transition metal dichalcogenides with either 1- or 2 layers per superlattice period ($n$=1, 2) and MX is a distorted rocksalt with 1 layer per period ($m$=1). As lattices are incommensurate and constituents are stacked along van der Waals surfaces, MLCs provide potential to realize novel material combinations without the physical constraints of traditional epitaxy.[10,11] Unfortunately the thermodynamic nature of products realized by high-temperature MLC synthesis precludes investigation of a full series of variably-stacked materials.

Kinetically-controlled, self-assembled misfit layer compounds remove the thermodynamic constraints which limit $m$ and $n$ in such systems. This development allowed study of systematic changes in structure and properties as a function of precise changes to nano-architecture via modulation of the stacking sequences (i.e., $m$, $n$ = 1, 2, 3, 4, …) and strategic control of defects and dopants. For example, charge density wave transitions and interlayer charge exchange were systematically investigated in series of experiments on [(MSe)$_{1+\delta}$]$_m$(VSe$_2$)$_n$ (M=Sn, Pb, Bi)[12–16] and [(MSe)$_{1+\delta}$]$_m$(TiSe$_2$)$_n$ (M=Sn, Pb, Bi),[17–21] respectively. Although the majority of known MLCs are in the sulfide space,[22] to date no such kinetically-controlled sulfide MLCs have been prepared, being instead widely distributed across the selenide, and, less frequently, the telluride chemical spaces.[22–24] Among others, the high vapor pressure and corrosive nature of sulfur presents a number of experimental challenges during the high vacuum deposition process typically used to prepare kinetically controlled films.

We previously reported a framework for the preparation of amorphous sulfide precursor films with controlled layer thicknesses and compositions generated by RF sputtering, a technique which mitigates the worst of the issues commonly seen in sulfide-based depositions.[25] Here, we expand on that work and introduce an independent sulfur source to create a series of crystalline [(SnS)$_{1+\delta}$]$_m$(TaS$_2$)$_1$ nanoscale superlattice compounds from specifically designed precursors, the first report of artificial sulfide MLC heterostructures with kinetically-controlled architectures. Through x-ray techniques and electron microscopy we explore the structure and composition of these films both globally and locally and confirm the formation of ordered superlattice nanocomposites programmed by periodic modulation of precursor constituents. We then implement an intentional grading approach to identify multiple stacking sequences across the surface of a single sample; this enables rapid identification of various structures and is a potential route toward the structural fine-tuning of material properties in sulfide superlattices.

**RESULTS AND DISCUSSION**

**Structure and composition of optimized $(SnS)_3(TaS_2)_1$ superlattice.** We choose the $[(SnS)_{1+\delta}]_m(TaS_2)_n$ system due to its established presence in MLC literature, where it has been synthesized with an $n=m=1$ layer configuration and a misfit parameter of 1.15.[4,26] Basal planes of constituent layers from the refined $[(SnS)_{1.15}]_1(TaS_2)_1$ structure are shown in **Figure 1a.** Reported MLCs show both SnS and $TaS_2$ with an orthorhombic crystal structure having space groups C2mb and C2mm, respectively, interleaved at monolayer length scales (crystalline assembly in **Figure 1b**).[27] Superconductivity was recently observed in single crystal samples of $(SnS)_1(TaS_2)_1$, further highlighting this system as warranting continued study.[4]

To create kinetically stabilized $[(SnS)_{1+\delta}]_m(TaS_2)_n$, layered thin film precursors are prepared by RF sputtering. Cations are sourced from an elemental tantalum target and a compound SnS target; a heated RF sulfur source is present as an additional source of reactive sulfur. The sputter rate from elemental Sn was too high to facilitate the smooth, thin, amorphous layers required in the precursor. A chamber diagram showing the relative positions of sources to the substrate is presented in **Figure 1c**. The shutters above each sputter plume are sequentially activated such that precursor layers contain the precise number of atoms to nucleate monolayers. A simplified schematic showing the thermally-induced self-assembly of a calibrated precursor into a crystalline heterostructure is shown in **Figure 1b**. Further experimental details can be found in the Methods section below. Signals from quartz crystal microbalances placed above each cation source during preparation of a [(SnS)$_{1+\delta}$]$_3$(TaS$_2$)$_1$ precursor film are shown alongside a schematic of the final heterostructure in **Figure 1d**. Note that in this work we use identical chemical nomenclature for both precursor films and products; this reflects the programmed design of the precursor film, which is confirmed by the experimental results discussed here.

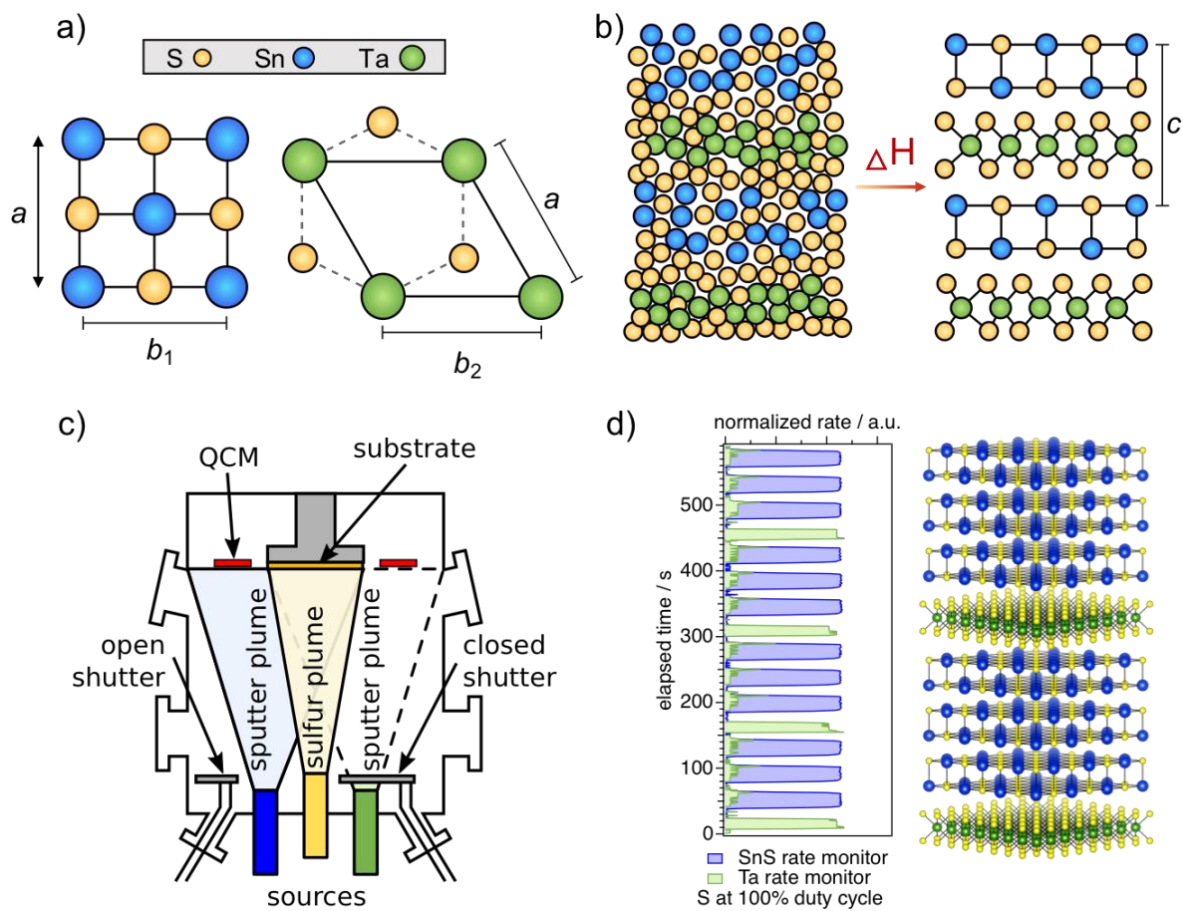

**Figure 1**. Synthesis of optimized [(SnS)$_{1+\delta}$]$_3$(TaS$_2$)$_1$. a) Basal planes for SnS and TaS$_2$ projected along the c-axis. b) Schematic of proposed state change between a precursor film (left) and a crystalline superlattice after annealing (right) in an example [(SnS)$_{1+\delta}$]$_1$(TaS$_2$)$_1$ structure. c) Side-view of sputter chamber geometry used to deposit precursor films. Orthogonal shuttered sputter sources (red and blue) and an unshuttered elemental sulfur cracker are used as sources; rates are monitored by quartz crystal microbalances (yellow squares). d) Experimental QCM signal for shutter cycling used to generate a [(SnS)$_{1+\delta}$]$_3$(TaS$_2$)$_1$ structure, shown as a model on right.

Composition and local structure measurements determined for an optimized a [(SnS)$_{1+\delta}$]$_3$(TaS$_2$)$_1$ by XRD, TEM, and EDS confirm a layered nanocomposite following the structure expected by XRD. These results are shown in **Figure 2**. **Figure 2a** shows specular XRD from the same precursor after annealing. The family of (00$l$) peaks arise from planes of atoms in the stacking direction of the superlattice. Whole-pattern fitting is used to model the experimental data and extract a superlattice period of 23.116 A. This value agrees well with published lattice parameters assuming three layers of SnS, one layer of TaS$_2$, and known interlayer spacings reported in the MLC system (~2-3 Å).[28] Symmetry of the constituent layers is observed by in-plane XRD, shown in **Figure 2b**, with peaks corresponding to a superposition of (*hk0*) planes of crystalline SnS and TaS$_2$. SnS is indexed to the cubic space group Pbcm with a = 0.410 nm, in close agreement with established values. The remaining/overlapping peaks index well to TaS$_2$ with a lattice parameter of 0.333 nm, but in-plane measurements cannot distinguish between the 1T and 2H polytypes since monolayers of each coordination type have identical projections onto the basal plane. From these fits a misfit parameter of 1.14 is calculated, in good agreement with MLC literature.

Planes of atoms with well-established ordering over several nanometers are clearly visible in the atomic resolution Bright field STEM micrograph collected from [(SnS)$_{1+\delta}$]$_3$(TaS$_2$)$_1$ (**Figure**

**2c**). Bright points appearing in some layers arise from columns of atoms that are on zone-axis. Rotational misregistration is observed both from different grains within a layer and across interfaces. Such turbostratic disorder is a common feature of misfit compound superlattices crystallized from designed precursors and attributed to a combination of weak van der Waals interactions between incommensurate layers and the kinetically-mediated crystallization process.[22,23] STEM-EDS spectral imaging as shown in **Figure 2d** shows defined, periodically-alternating modulation of the different cations throughout the thickness of the film. While a number of stacking faults and exfoliation-like behavior is visible in the Bright field STEM image, overall order of the superlattice structure is maintained despite a global compositional offset from the composition of the heterostructure itself as determined by in-plane XRD (**Figure 2b**) (see discussion below); this may suggest that the SnS-TaS$_2$ superlattice is more robust than some of its selenide counterparts in maintaining a distinct, thermodynamically unachieved architecture.[15]

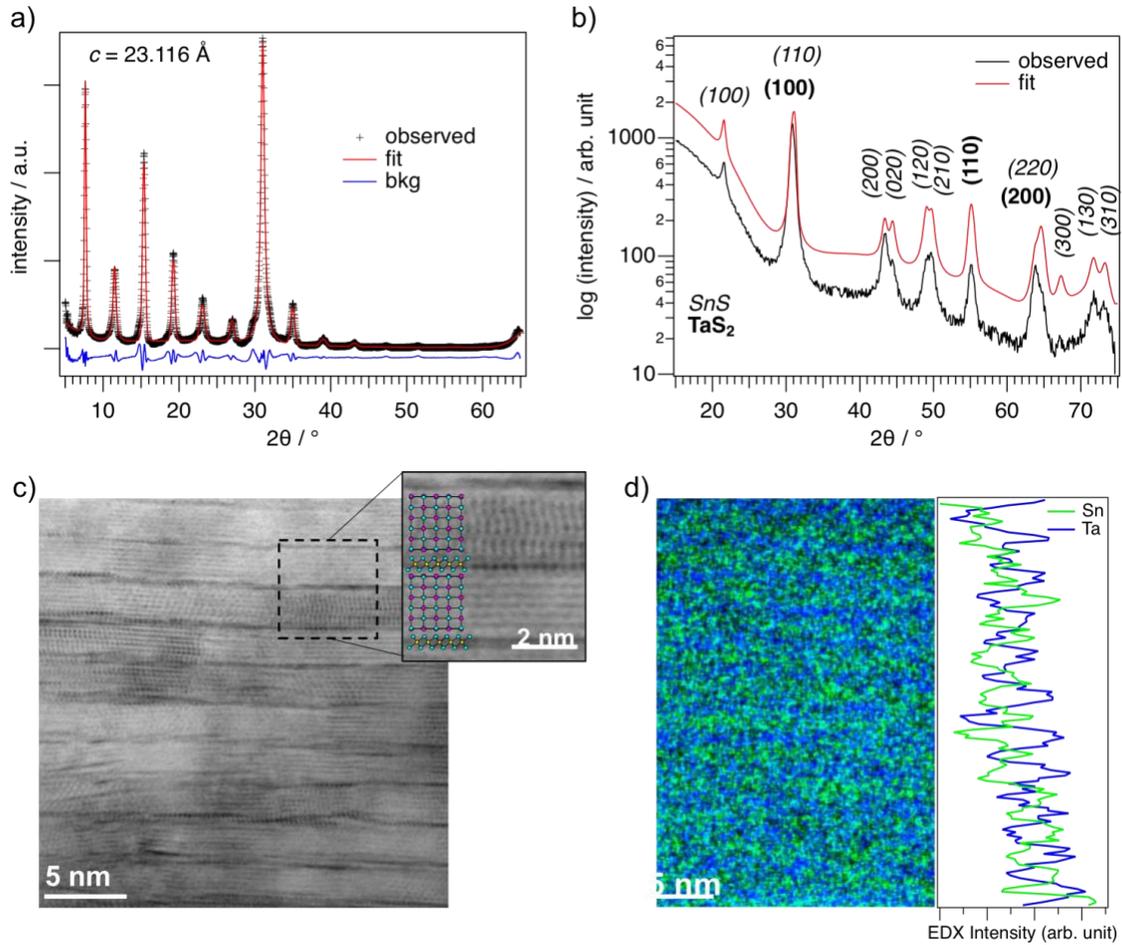

**Figure 2.** a) Fit of specular XRD pattern for an optimized [(SnS)$_{1+\delta}$]$_3$(TaS$_2$)$_1$ compound in which the heterostructure repeating unit along the c-axis is calculated. b) Fit of an in-plane XRD pattern for the same structure, showing peaks corresponding to constituent TaS$_2$ and SnS layers. c) Representative bright field scanning transmission electron micrograph of crystalline [(SnS)$_{1+\delta}$]$_3$(TaS$_2$)$_1$ film. Contrasting atom brightness highlights the difference between Sn- and Ta-containing layers, as well as showing evidence of turbostratic disorder between layers of like phase. The corresponding crystal structure is shown in inset. d) STEM-EDS spectral imaging of [(SnS)$_{1+\delta}$]$_3$(TaS$_2$)$_1$ film shows periodic alternation of Ta and Sn atoms in the out-of-plane direction, consistent with the structure predicted by XRD and TEM.

**Structural tunability via programmatic precursor changes.** Once conditions resulting in high-quality $[(SnS)_{1+\delta}]_3(TaS_2)_1$ were established, the chamber shutters were reprogrammed to target a series of $[(SnS)_{1+\delta}]_m(TaS_2)_1$ superlattices with *m* = 3, 4, and 5. The structure of films resulting from various annealing conditions were determined by diffraction; patterns are shown in **Figure 3**. A schematic of these stacking sequences is presented in **Figure 3a** alongside X-ray reflectivity (XRR) patterns showing annealed films to be ultra-smooth, with Kiessig fringes extending beyond 5° 2θ. The low-angle Bragg peaks seen in XRR are consistent with the large superlattice periods expected from each precursor. The repeat unit thickness determined by specular XRD shown in **Figure 3b** are calculated as 2.31 nm, 2.86 nm and 3.37 nm for $[(SnS)_{1+\delta}]_3(TaS_2)_1$, $[(SnS)_{1+\delta}]_4(TaS_2)_1$, and $[(SnS)_{1+\delta}]_5(TaS_2)_1$, respectively. These values are again in good agreement with the lattice parameter calculated for the proposed stacking sequences. A linear fit of these thicknesses against the integral number of SnS layers per repeating unit indicates the $TaS_2$ monolayers provide 0.67 nm of thickness to the repeating unit (intercept), while each additional SnS layer increases repeat unit thickness by 0.55 nm (slope). The slight deviations from SnS and $TaS_2$ is reasonable because the interface between constituents is distinct from that between layers in the bulk compounds.

RBS-calibrated XRF measurements of thin films, following the method in Ref [29], show that significant excess Sn is present in all cases. We calculate a Sn:Ta ratio of 6:1 in the $[(SnS)_{1+\delta}]_3(TaS_2)_1$ film, whereas the ratio required to accommodate structural misfit is 3.5:1. Film composition does not change appreciably upon anneal. Unsurprisingly, this excess Sn manifests itself structurally; specular XRD patterns shows a shoulder around 33° 2θ, corresponding to the (*400*) plane of cubic SnS.[30] These global composition measurements are complemented by in-plane diffraction patterns shown in **Figure 3c**, where patterns for the three stacking architectures are normalized to the lone TaS$_2$-specific peak with diffraction from the (110) planes. Note that the signal to noise ratio increases with decreasing SnS layers due to this scaling. SnS-specific planes shows an increase in intensity for increasing repeat unit thickness, as there is a concomitant decrease in the relative fraction of TaS$_2$. The decrease in crystal quality with increasing *m* might be due to several factors, such as roughening of thicker SnS blocks or imperfect calibration of shutter timing during both cation deposition and sulfurization. Attempts to grow films with *m*<3 were unsuccessful, possibly due to the size limit of SnS adsorbates during precursor growth. Full diffraction patterns can be seen in the SI.

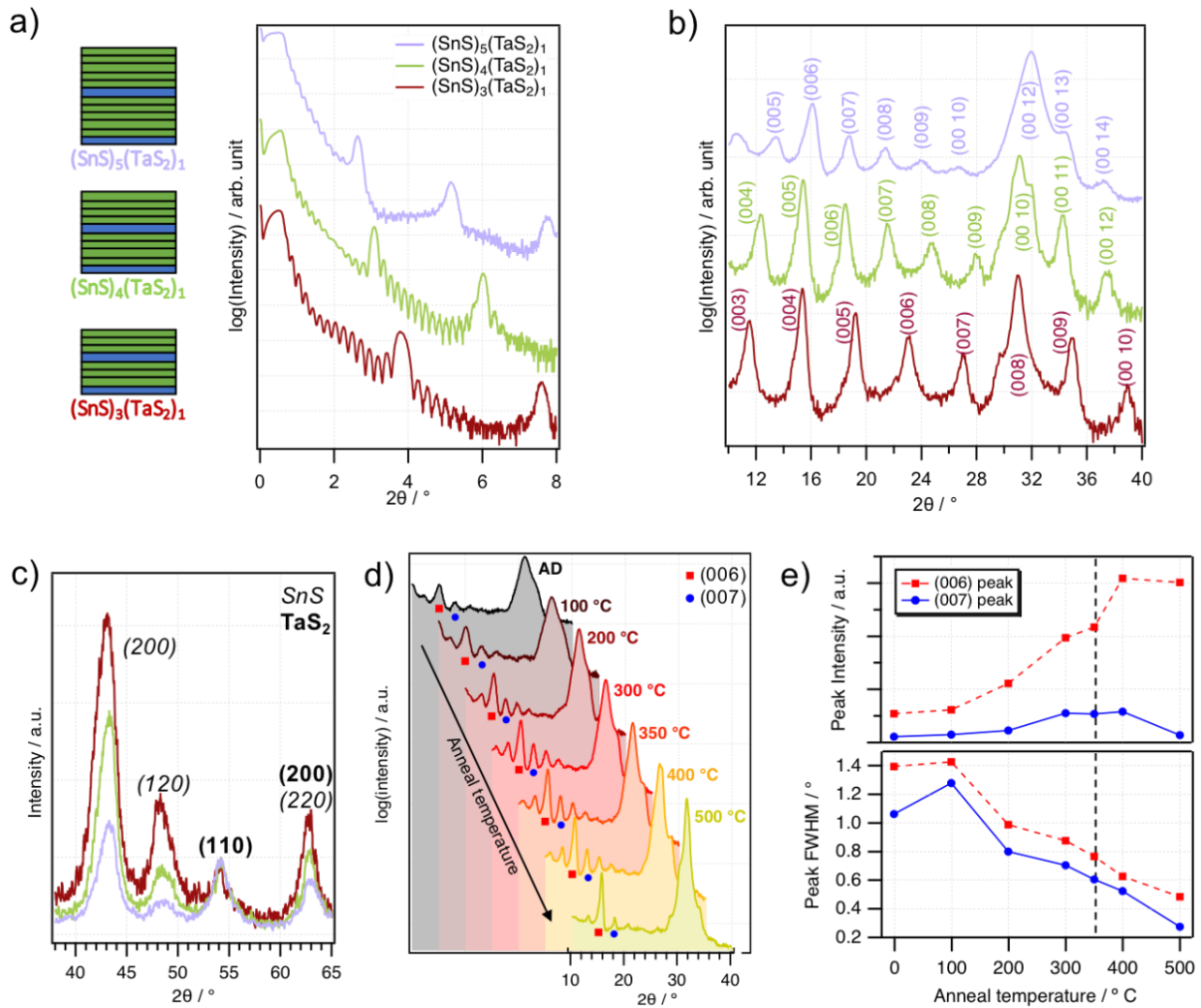

**Figure 3**. a) Reflectivity and b) specular x-ray diffraction of $[(SnS)_{1+\delta}]_m(TaS_2)_1$ heterostructures, where m = 3, 4, or 5. A schematic of the repeating unit for each heterostructure architecture is show adjacent to XRR patterns. c) In-plane XRD shows presence of peaks associated with SnS and $TaS_2$, confirming these compounds as constituents in the superlattice. Patterns are normalized to height of $TaS_2$ (110) plane to show relative changes in SnS. d) XRD patterns for a precursor film with a $[(SnS)_{1+\delta}]_5(TaS_2)_1$ structure annealed at various temperatures. e) Peak intensity (top) and full width at half max (bottom) for (006) and (007) peaks corresponding to the peaks notated in d).

The [(SnS)$_{1+\delta}$]$_5$(TaS$_2$)$_1$ sample was used to investigate formation and thermal degradation of the layered structure. Precursor films were annealed for 20 minutes at temperatures between 100 °C and 500 °C. Superlattice diffraction presented as a heatmap as a function of annealing temperature and scattering angle is presented in **Figure 3d**. The optimal condition balances in-plane crystallization of the superlattice against cross-plane diffusion and/or sulfur effusion leading to breakdown of the metastable heterostructure. Specific to the present study, annealing conditions must also balance crystallization of the superlattice against diffusion and agglomeration of excess Sn (T$_{melting}$=232°C), that may serve to increase roughness, inhomogeneity, and stacking defects in the film.

We analyze the quality of patterns in **Figure 3d** by calculating the intensity and full width at half max (FWHM) for the even (006) and odd (007) diffraction peaks (**Figure 3e**). Whereas even (*00l*) peaks can overlap with phase-segregated constituents, odd (*00l*) peaks arise solely from superlattice ordering. While overall film crystallinity appears to improve up to 400 °C, anneal temperatures above 350 °C show a breakdown in diffraction intensity for the superlattice-specific (007) peak likely associated with excess cross-plane diffusive mobility. This analysis suggests an optimal annealing temperature near 350 °C to maintain integrity in the FWHM of superlattice peaks while avoiding the increasingly dominant presence of the overlapping SnS *(400)* peak around 31° 2θ at higher temperatures. Annealing time was then varied to fully optimize for the annealing condition that produces the best crystalline superlattice. These data indicate an optimal annealing of 10 minutes. XRD patterns corresponding to annealing conditions in which time are varied are shown in the Supporting Information.

**Rapid synthesis of multiple architectures.** Intentional grading afforded by the geometric design of our sputter chamber allows for rapid exploration of multiple architectures across a single film. Such combinatorial approaches have proven effective for rapidly surveying stability and properties across chemical spaces[31,32] and have even been applied to metastable chalcogenides[33,34] but to the best of our knowledge these approaches have not been utilized for preparation of nanolayered heterostructures. As described in our previous work, precursor films can be prepared such that thickness gradients of SnS and Ta are orthogonal to one another on a sample surface.[25] The sulfur flux is nominally identical across the library under the assumption that sulfurization is self-limiting due to the high vapor pressure. Compositional and structural characterization of a sample prepared in this way is shown in **Figure 4**, showing systematic changes across the sample surface. Relative intensities of cation counts from x-ray fluorescence are plotted in **Figure 4a**, showing an increasing fraction of SnS moving up the rows of the sample. The gradient along the Ta direction is much shallower than that in the SnS direction, as evidenced by spatially distinct lattice parameter changes in the as-deposited film (see SI). This is a result of sample position and relative sputter rates/orientation of the two sources.

**Figure 4b** shows the calculated repeat unit thickness for points across the sample surface before and after annealing. Here, Point 1 corresponds to the top left corner of the sample and Point 25 corresponds to bottom right. While the as-deposited sample shows repeat unit thickness changing systematically across rows, a unique stacking sequence is not found in an intermediate row of the annealed sample (**Figure 4c** inset, second from bottom) and instead a lower quality $[(SnS)_{1+\delta}]_5(TaS_2)_1$ stacking structure was observed. We suggest that the local composition of the precursor film in this region was too imprecise to nucleate a superlattice with an integer number of monolayers and so instead compensated by forming the closest available stacking structure. A comparison of these diffraction patterns is available in the Supporting Information.

While the analysis provided here is not sufficient to conclusively determine the precise stacking sequence, given our experimental setup and the calculated repeat unit thicknesses we argue that stacking sequence changes occur from insertion of additional SnS layers and span a range between $[(SnS)_{1+\delta}]_4(TaS_2)_1$ and $[(SnS)_{1+\delta}]_7(TaS_2)_1$. Experimentally-determined repeat unit thicknesses for each of the regions identified is compared to the calculated thickness for the proposed nanoarchitecture in **Figure 4c**, showing excellent agreement. Superlattice diffraction patterns from these points are presented in **Figure 4d**. Colored boxes in the inset of **Figure 4c** indicate positions where diffraction was measured from crystalline $[(SnS)_{1+\delta}]_m(TaS_2)_n$ superlattices with systematic increases to the number of layers.

Despite the reduced quality of superlattice peaks relative to that in the fully optimized sample in **Figure 1**, this method demonstrates clear proficiency at rapid identification of a variety of heterostructure architectures. We note that the entire surface of the sample was exposed to the same shutter cycling protocol yet produced distinct stacking sequences; this is likely only possible due to the compound SnS target and/or co-exposure of the substrate to the sputter plume and sulfur vapor. This indicates that precise layer modulation may be less critical here than by methods where cations/anions are also modulated in the precursor. Further study is required to parse out details of precursor tolerance and heterostructure formation mechanisms.

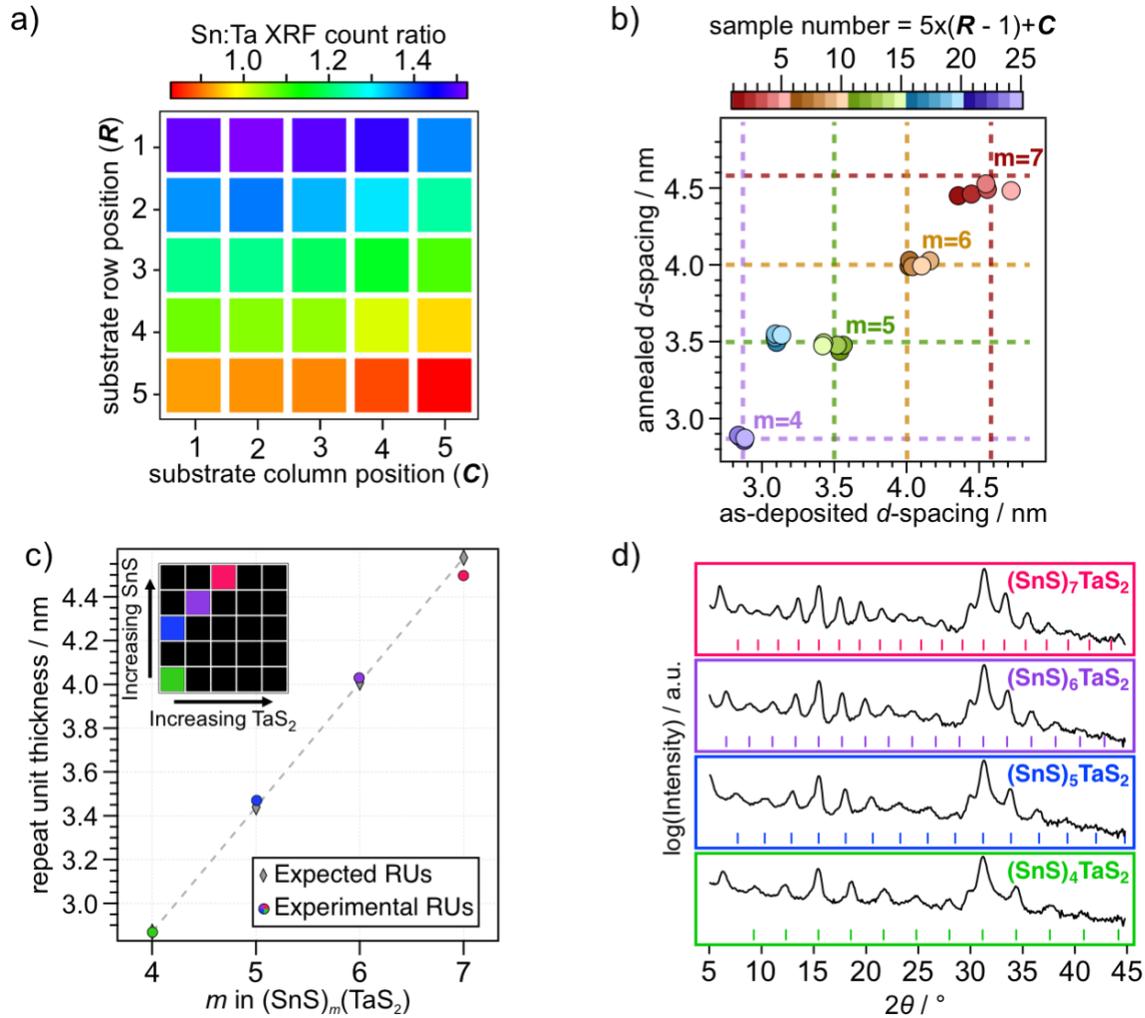

**Figure 4.** a) Composition heat map of relative cation counts determined by XRF. The gradient along the SnS direction is far steeper than that along the $TaS_2$ direction. b) Plot of calculated repeat unit thickness for points on sample before and after annealing. Dashed lines reference location of repeat unit for ideal $(SnS)_m(TaS_2)_1$ compounds. c) Inset shows spatial schematic of points on a sample with intentionally graded thickness and composition. Repeat unit thickness of colored points in inset, corresponding to highlighted regions of sample with different stacking sequences. The calculated repeat unit thickness for each $[(SnS)_{1+\delta}]_m(TaS_2)$ structures is denoted in grey. d) Specular x-ray diffraction of the points denoted in a) and b), showing heterostructures self-assembled with different stacking sequences across a single sample.

## CONCLUSIONS

In this work we successfully utilize designed thin film precursors to prepare and characterize sulfide-based layered nanoscale superlattices of similar structure and composition kinetically prepared selenide structures. We demonstrate control of the nanoarchitecture by targeted synthesis of $[(SnS)_{1+\delta}]_m(TaS_2)$ heterostructures ($3 \leq m \leq 5$). This is the first demonstration of such kinetically-controlled stacking sequences of a sulfide-based misfit layer compound. We further demonstrate that rapid exploration of various nanoarchitectures is possible through preparation of intentionally graded precursors which span a range of thickness and composition ratios between constituent compounds. Using this approach we synthesize $[(SnS)_{1+\delta}]_m(TaS_2)$ heterostructures ($4 \leq m \leq 7$) from a single precursor film. This work opens up exploration of structure-controlled-modulation of optical and electronic properties (i.e. CDW transitions) by designed layer sequencing in the large chemical space of 2D sulfide materials.

## METHODS

**Deposition of precursor films.** Precursor films are deposited in an RF sputter chamber using a system and calibration procedure described in our previous work, with some modification.[25] An elemental Ta and compound SnS target are used as a source of cations in the film. Supplemental sulfur is provided by an RF solids cracker from Oxford Applied Research. Elemental sulfur is evaporated in a Knudsen cell at the base of the cracker and transported by flowing argon through an RF coil, which generates a sulfur/argon plasma. Unless otherwise specified, argon flow is 6 sccm, the Knudsen cell is set to 60 C, and the RF coil operates at 30 W. The cracker is un-shuttered and sulfur is constantly depositing either alongside the target material or as an isolated layer when both target shutters are closed. The deposition rate is determined for various sputter scenarios is determined. The deposition rate of (sulfur + SnS) and (sulfur + Ta) is 0.009 and 0.045 nm/s, respectively; when both metallic sources are closed, the deposition rate of elemental sulfur on a layer of SnS or Ta is 0.004 and 0.003 nm/s respectively. Overall chamber pressure was held constant at 2.8 mTorr by adjusting the throttle position at a turbomolecular pump.

**Annealing and structural characterization.** Calibrated precursor films were crystallized by heating on a hotplate in an inert environment. Structural characterization of layering in amorphous and crystallized films was primarily done via x-ray reflectivity (XRR) and x-ray diffraction (XRD) to determine thicknesses of constituent repeating units as-deposited and as a function of thickness. Structural properties were obtained using a Rigaku Smartlab diffractometer using a 5 mm height minimizing slit and a 0.01 deg 2θ step size. An understanding of constituent layer structure was obtained by performing in-plane XRD on a Rigaku Ultima IV using a thin film geometry and a 0.01 deg 2θ step size.

**Composition analysis.** Composition measurements were performed using Rutherford backscattering spectrometry (RBS) and x-ray fluorescence (XRF). RBS measurements were performed using the instrument model, parameter, and analysis procedures described previously.[25] Vacuum XRF measurements were performed with a Bruker M4 Tornado. Counts were detected using two Si drift detectors. Using a protocol similar to that described in Reference [29], spectra were then processed to generate a correlation between XRF line intensity and RBS measurements. Background scans of the silicon substrates were subtracted from the spectra generated by the thin film; peak intensity was integrated about the characteristic x-ray line for elements of interest. The Kα x-ray energy at 2.309 keV used for sulfur and the Lα x-ray energies at 8.146 and 3.444 keV were used for Ta and Sn, respectively.[35] Data was processed using a custom Igor routine.[36]

**STEM and EDS analysis.** Cross-sectional specimens of the as-deposited and annealed [(SnS)$_{1+\delta}$]$_3$(TaS$_2$) precursors were prepared using standard focused ion beam methods and in-situ specimen manipulation to Cu grids. Specimens were prepared using an FEI Helios 600i DualBeam focused ion beam/scanning electron microscope (FIB/SEM) with a Ga ion source, initially at 30 kV accelerating voltage. A final 2 kV FIB cleaning procedure largely removed Ga surface damage and reduced the specimen thickness to <50 nm. (S)TEM imaging and EDS spectral imaging of the specimens were acquired using an FEI Titan 80-300 probe corrected and monochromated (S)TEM operating at 300 keV.

AUTHOR INFORMATION

**Corresponding Author**

*Dr. Sage Bauers

 sage.bauers@nrel.gov

**Present Addresses**

†Materials Measurement Laboratory, National Institute of Standards and Technology, Boulder CO 80305.

**Author Contributions**

The manuscript was written through contributions of all authors.

**Funding Sources**

   This work was authored by the National Renewable Energy Laboratory (NREL), operated by Alliance for Sustainable Energy, LLC, for the U.S. Department of Energy (DOE) under Contract

No. DE-AC36-08GO28308. D.M.R., J.D.P., S.R.B, and A.Z. acknowledge funding provided by the Laboratory Directed Research and Development (LDRD) Program at NREL for the growth and characterization of all materials. The views expressed in the article do not necessarily represent the views of the DOE or the U.S. Government.

ABBREVIATIONS
XRD x-ray diffraction, TEM transmission electron microscopy, STEM scanning transmission electron microscopy, EDS energy dispersive spectroscopy, XRF x-ray fluorescence, RBS Rutherford backscattering spectrometry